\documentclass{article}  
          \begin{document}
           \title{Use of Mathematical Logical Concepts in Quantum Mechanics: An Example}
          \author{Paul Benioff\\
           Physics Division, Argonne National Laboratory \\
           Argonne, IL 60439 \\
           e-mail: pbenioff@anl.gov}
           \date{\today}

          \maketitle
          \begin{abstract}
            The representation of numbers by product states in quantum mechanics
            can be extended to the representation of words and
            word sequences in languages by product states.  This
            can be used to study quantum systems that generate text that has meaning.
            A simple example of such a system, based on an example described by
            Smullyan, is studied here. Based on a path interpretation for some word
            states, definitions of truth, validity, consistency
            and completeness are given and their properties
            studied. It is also shown by means of examples that the relation between
            the potential meaning, if any, of word states and the
            quantum algorithmic complexity of the process
            generating the word states must be quite complex or nonexistent.
          \end{abstract}

\section{Introduction}
Quantum computers have been much studied in recent years due
mainly to the possibility of solving some important problems more
efficiently on quantum computers than is possible on classical
machines \cite{Shor,Grover}.  However quantum computers and
quantum robots \cite{BenioffQR} are also interesting from other
viewpoints.  For example, a study of these systems may help to
determine what properties a  quantum system must have to conclude
that it has significant characteristics of intelligence. If
quantum mechanics is universally applicable, then many
intelligent quantum systems exist (e.g. the readers of this
paper). The fact that these systems are macroscopic, which may be
necessary, does not contradict the fact that they are also
quantum systems.

Another aspect originates in the fact that basis states of
multiqubit systems in quantum computers are product states
$|\underline{S}\rangle$ of the form $|\underline{S}\rangle =
\otimes_{j=1}^{n}|\underline{S}_{j}\rangle$ where $ \underline{S}$
is a function from $\{1,\cdots ,n\}$ to $\{0,1\}$. and
$|\underline{S}_{j}\rangle$ is the state of the jth qubit. Here
the state $|\underline{S}\rangle$ is supposed to  be a binary
representation of a nonnegative integer. Even though this
representation is assumed implicitly in the literature, it is not
trivial, especially regarding conditions that a composite physical
system must satisfy in order that it admits states representing
numbers \cite{BenioffRNQM}.

The notion that states $|\underline{S}\rangle$ represent the
nonnegative integers can be extended to $k-ary$ representations of
more general types of numbers, such as all integers and rational
numbers \cite{BenioffALG}. The representations can also be
extended by considering the states $|\underline{S}_{j}\rangle$ to
represent symbols  in some language. The language can be formal as
is the case for axiom systems studied in mathematical logic, or
informal as is the case in English.  In this case the symbol basis
states would include orthogonal states for each  of the 26 letters
in the alphabet and states for some  punctuation symbols. Word
states would consists of products of the symbol basis states
excluding the spacer symbol state.  These would be used to
separate the different words.

These representations are of interest for several reasons.  As
part of an attempt to characterize intelligent quantum systems,
one wants to understand what physical properties a quantum system
must have so that it can be said to be creating text that has
meaning to the system generating the text.  Another is related to
the need to develop a coherent theory of mathematics and physics
together that is maximally internally self consistent. In such a
theory  one would expect  mathematical logical concepts to be
closely integrated with quantum mechanics or some generalization
such as quantum field theory. Since mathematical logic deals with
systems of axioms as words in a language and their
interpretation, such an integration would require the
representation of words by quantum states.

The potential importance of this has been recognized by other
work. Included are recent work on theories of everything
\cite{Tegmark},  an attempt to use Feynman diagrams to represent
expressions in propositional logic \cite{Schmidhuber}, and other
relevant work \cite{DeEk,Spector}.

In order to see how mathematical logical concepts such as truth,
validity, consistency, and completeness might be used in quantum
mechanics, an example will be studied that is based on a
simplification of a simple machine described by Smullyan
\cite{Smullyan}.  The description of the simplified machine, given
in the next section, is followed by a description of a quantum
machine that is similar to a quantum Turing machine
\cite{Deutsch}. Additional details are given in \cite{BenDTVQM}.

\section{Smullyan's Machine}

The simplified version of Smullyan's machine M \cite{Smullyan},
used here prints, one symbol at a time,  a nonterminating string
of any one of the five symbols $P,\; \sim ,\; (,\; ),\; 0$. Words
are defined as any finite strings of symbols that exclude the $0$
which denotes a spacer symbol.  Based on this the machine prints a
steadily growing string of words separated by finite spacer
strings.

Some of the words, which are separated from other words by spacer
strings, are assigned a meaning.  These words, referred to as
sentences, are $P(X)$ and $\sim P(X)$ where $X$ is any word that
is not a sentence.  The strings $0P(\sim (PP)0$ and $0\sim
P()P)\sim ()0$ where $X= \sim (PP$ and $X=)P)\sim ($ are examples
of these two types of sentences with separating $0s$ shown. The
intended meaning of $P(X)$ is that $X$ is printable and that of
$\sim P(X)$ is that $X$ is not printable.

The restriction that $X$ is not a sentence is  not present in
Smullyan's original example.  It is made here to keep things
simple and avoid inference chains generated by words of the form
$P(P(X)),P(\sim P(X))$, etc.. Removal of the restriction is
discussed in \cite{BenDTVQM}.

The term  "printable" refers to the dynamical description of M. If
the dynamical description of M correctly predicts the behavior of
$M$, then any word that is printable will be printed sooner or
later.  If the word is not printable, then it will never be
printed.

Based on the assigned meaning of these words, $P(X)$ is defined to
be {\em true} if $X$ is printable. It is false if $X$ is not
printable. $\sim P(X)$ is {\em true} if $X$ is not printable and
false if $X$ is printable. These definitions relate truth and
falseness of the sentences to the dynamics $U$ of $M$. However
nothing is said so far about whether these statements are true or
false.

This is accounted for by defining the dynamics of M to be {\em
valid} if any printable sentence is true. It follows that false
sentences are not printable. It is also the case that the dynamics
is valid if no sentences are printable. This possibility is
avoided by requiring the dynamics to  be {\em complete}. That is,
it must be such that  for all $X$ that are not sentences, either
$P(X)$ or $\sim P(X)$ is printable. It is {\em consistent} if at
most one of $P(X)$ or $\sim P(X)$ is printable.

\section{Quantum Machine Model} \label{QMM}
Let M be a multistate quantum system or a head moving along a one
dimensional lattice of quantum systems at sites $1,2,\cdots$. The
basis states of the head M have the form $|\ell ,j\rangle$ where
$\ell$ is an internal head state label and $j$ the lattice
position of M. The Hilbert space of states associated with the
system at lattice site $j$ is spanned by a basis of five states
$|P,j\rangle,\; |\sim ,j \rangle,\; |(,j\rangle,\; |),j\rangle,
|0,j\rangle$. In what follows these states will be designated by
either $|S_{j}\rangle$ or as $|\underline{S}_{j}\rangle$. The
state $|\underline{S}_{j}\rangle \equiv
|\underline{S}(j),j\rangle$ refers to the symbol state of the
system at site $j$ as the value of a function $ \underline{S}$ at
$j$ where $ \underline{S}$ is a function from the set of lattice
sites to the set of five symbols. The state $|S_{j}\rangle\equiv
|S,j\rangle$ refers to the system at site $j$ in state $|S\rangle$
with no reference to a function.

The lattice basis states have the form $|\underline{S}\rangle =
\otimes_{j=1}^{\infty}|\underline{S}_{j}\rangle$ where at most a
finite number of the lattice systems are in states $|P\rangle
,|\sim\rangle, |(\rangle, |)\rangle$ different from $|0\rangle$.
The finiteness restriction means that
$|\underline{S}_{j}\rangle\neq |0\rangle$ for at most a finite
number of $j$ values. This restriction is imposed to keep the
Hilbert space spanned by the $|\underline{S}\rangle$ separable.

Let $|\underline{S}_{[a,b]}\rangle
=\otimes_{j=a}^{b}|\underline{S}_{j}\rangle$ denote the product
state for the symbol states in the lattice interval $a\leq j\leq
b$. There is an obvious map of $|\underline{S}_{[a,b]}\rangle$ to
a basis state
$|\underline{S}_{[a,b]}\rangle\otimes|\underline{0}_{\neq
[a,b]}\rangle$ which has $0s$ at all sites outside $[a,b]$. A word
state is defined to be any $|\underline{S}_{[a,b]}\rangle$ where
$|\underline{S}_{j}\rangle\neq |0_{j}\rangle$ for each $j$ in
$[a,b]$. States $|\underline{S}_{[a,b]}\rangle$ where
$|\underline{S}_{j}\rangle =|0_{j}\rangle$ for each $j$ in $[a,b]$
will be referred to as spacer string states or as empty word
states. Based on this any basis state $|\underline{S}\rangle$ is
clearly a finite sequence of alternating word and spacer string
states.

The dynamics of $M$ is  such that $M$ moves in one direction, one
site per step, and interacts with the lattice systems at and just
behind the location of $M$ and with no others. This choice of the
interaction range for $M$ is arbitrary and is done to to keep
things simple. This dynamics is described by a time step
operator\footnote{As described $U$ is not unitary as it moves the
head in one direction on a one directional infinite lattice.
Unitarity can be restored by defining the lattice as extending
from $-\infty$ to $\infty$. However this will not be done as it
adds nothing to the discussion.} $U$ for which all the nonzero
matrix elements have the form $\langle \ell^{\prime},j+1,
S_{j}^{\prime},S_{j-1}|U|\ell,j,S_{j}^{\prime \prime}
,S_{j-1}^{\prime}\rangle.$   Here $\langle \ell^{\prime},j+1,
S_{j}^{\prime},S_{j-1}|U|\ell,j,S_{j}^{\prime \prime}
,S_{j-1}^{\prime}\rangle$ gives the amplitude for M in state
$|\ell\rangle$ and at position $j$, and lattice systems at sites
$j$ and $j-1$ in symbol states $|S_{j}^{\prime\prime}\rangle$ and
$|S_{j-1}^{\prime}\rangle$, moving to site $j+1$ and changing to
state $|\ell^{\prime}\rangle$. The symbol states change to
$|S_{j}^{\prime}\rangle$ and $|S_{j-1}\rangle$.

At time $0$ the overall state of M and the lattice is given by
$|i,2,\underline{0}\rangle = \Psi(0)$ where
$|\underline{0}\rangle=\otimes_{j=1}^{\infty}|\underline{0}_{j}\rangle$
is the state denoting all lattice systems at sites $1,2,\cdots$ in
the spacer symbol state and $|i,2\rangle$ show $M$ in initial
state $|i\rangle$ and at location $2$.  At time step $n$ the
system is in state $\Psi(n)=U^{n}\Psi(0)$. This can be expressed
as a Feynman sum over symbol string  states as
\begin{eqnarray}
\Psi(n)=|\underline{0}_{[>n+1]}\rangle\otimes\sum_{\ell_{n},S_{n+1}^{\prime}}\sum_{
\underline{S}_{[1,n]}} |\ell_{n},n+2,
S_{n+1}^{\prime},\underline{S}_{[1,n]}\rangle  \nonumber \\
\langle\ell^{\prime},n+2,S_{n+1}^{\prime},\underline{S}_{[1,n]}
|U^{n} |i,2,\underline{0}_{[1,n+1]}\rangle \label{feyps}
\end{eqnarray}
The sum $\sum_{ \underline{S}_{[1,n]}}$ is over all $5^{n}$ length
$n$ symbol string states (including the spacer) of lattice systems
at sites $1,\cdots,n$. The sums $\sum_{\ell_{n},S_{n+1}^{\prime}}$
are over all $M$ states and over all five symbol states for the
site $n+1$. The states $|\underline{0}_{[>n+1]}\rangle$ and
$|\underline{0}_{[1.n+1]}\rangle$ are the constant $0$ states at
all lattice positions $>n+1$ and at lattice positions $1,\cdots
,n+1$ respectively. The separation of the state of the first $n$
systems from the $n+1st$ in the sums is based on the fact that in
future time steps $(>n)$ M is at positions $>n+1$ and no longer
interacts with the lattice systems at sites $1,\cdots ,n$.

To obtain Eq. \ref{feyps} one first notes that
$U^{n}|i,2,\underline{0}\rangle =
|\underline{0}_{[>n+1]}\rangle\otimes
U^{n}|i,2,\underline{0}_{[1,n+1]}\rangle$ as $U^{n}|i,2\rangle$
does not interact with lattice systems at sites $>n+1$. Expansion
between the $U$ operators in a complete set of states (where the
summations are understood) gives
\begin{eqnarray}
U^{n}|i,2,\underline{0}_{[1,n+1]}\rangle  =
U^{n-1}|\ell_{1},3,S_{2}^{\prime},S_{1},\underline{0}_{[3,n+1]}\rangle
\times \nonumber \\
\langle\ell_{1},3,S_{2}^{\prime},S_{1}|U|i,2,0_{2},0_{1}\rangle
 = U^{n-2}|\ell_{2},4,S_{3}^{\prime},S_{2},S_{1},
\underline{0}_{[4,n+1]}\rangle\times \nonumber \\ \langle
\ell_{2},4,S_{3}^{\prime},S_{2}|U|\ell_{1},3,0_{3},S_{2}^{\prime}\rangle\langle
\ell_{1},3,S_{2}^{\prime},S_{1}|U|i,2,0_{2},0_{1}\rangle = \cdots \nonumber \\
 = |\ell_{n},n+2,S_{n+1}^{\prime},\underline{S}_{[1,n]}\rangle
\langle \ell_{n},n+2,S_{n+1}^{\prime}, S_{n}|U|\ell_{n-1},n+1,
0_{n+1},S_{n}^{\prime}\rangle\times \nonumber \\ \cdots   \langle
\ell_{2},4,S_{3}^{\prime},S_{2}|U|\ell_{1},3,0_{3},S_{2}^{\prime}\rangle\langle
\ell_{1},3,S_{2}^{\prime},S_{1}|U|i,2,0_{2},0_{1}\rangle.
 \label{feypsa}
\end{eqnarray}

Here each state $|S_{j}\rangle$, created by the action of $U$ on
the state $|S^{\prime}_{j}\rangle$ with M at site $j+1$, is
passed to the left with no change through the successive $U$
operators. This occurs because $M$ is at lattice sites $>j+1$
where it does not interact again with a system at site $j$.
Carrying out all the intermediate $\ell$ and $S^{\prime}$ sums by
use of the completeness relations gives
\begin{eqnarray}
U^{n}|i,2,\underline{0}_{[1,n+1]}\rangle =\sum_{
\underline{S}_{[1,n]}}
\sum_{\ell_{n},S_{n+1}^{\prime}}|\ell_{n},n+2,S_{n+1}^{\prime},
\underline{S}_{[1,n]}\rangle\times \nonumber \\  \langle
\ell_{n},n+2,S_{n+1}^{\prime},
\underline{S}_{[1,n]}|U^{n}|i,2,\underline{0}_{[1,n+1]}\rangle
\label{feypsb}
\end{eqnarray} which gives Eq. \ref{feyps}.

From the definition of $|\underline{S}_{[1,n]}\rangle$, which
includes spacer symbol states, one sees that each state
$|\underline{S}_{[1,n]}\rangle$ is a product of word states
separated by spacer string or empty word states.  If $t$ is the
number of alternating empty and nonempty word states in
$|\underline{S}_{[1,n]}\rangle$, then $1\leq t\leq n$. For each
value of $t$ $|\underline{S}_{[1,n]}\rangle$ can be written as
\begin{equation} |\underline{S}_{[1,n]}\rangle =
|\underline{X}^{h_{t}}_{\nu(t)}\rangle \otimes
|\underline{X}^{h_{t-1}}_{\nu(t-1)}\rangle\otimes \cdots \otimes
|\underline{X}^{h_{1}}_{\nu(1)}\rangle.
\label{wdstrng}\end{equation} Here $\nu (j)$ is a two valued
function with values $0$ or $1$ with the property that the values
alternate.  That is, $\nu (j+1)=1-\nu (j)$.

If $\nu (j) =0$ then the state
$|\underline{X}^{h_{j}}_{\nu(j)}\rangle$ is a spacer string state
of length $h_{j}$. If $\nu (j) =1$ then
$|\underline{X}^{h_{j}}_{\nu(j)}\rangle$ is a word state of length
$h_{j}$. Since $|\underline{S}_{[1,n]}\rangle$ is a product of $n$
symbol states, the $h_{j}$ must satisfy $\sum_{j=1}^{t}h_{j} =n$.
Each of the $t$ states has at least one symbol state, so $1\leq
h_{j}$ for $j=1,\cdots ,n$.

Based on this each state $|\underline{S}_{[1,n]}\rangle$ can be
written as a word string or word path state.  There are four
possibilities depending on whether $t$ is even or odd and $\nu(1)
=0$ or $\nu(1)=1$. If $t$ is even and $\nu(1) =0$ then
$|\underline{X}^{h_{t}}_{\nu(t)}\rangle$ is a word state and
$|\underline{X}^{h_{1}}_{\nu(1)}\rangle$ is a spacer string state.
The other three possibilities refer to the other three
possibilities of initial and final states in Eq. \ref{wdstrng}.

Based on Eq. \ref{wdstrng} one sees that for each value of $t$,
$|\underline{S}_{[1,n]}\rangle$ is a word path state
$|\underline{p}\rangle$ with $t$ (empty and nonempty) words with
$|\underline{p}(j)\rangle =
|\underline{X}^{h_{j}}_{\nu(j)}\rangle$. This can be used to
expand $\Psi(n)$ as a sum over word path states in a form similar
to the sum over symbol string states shown in Eq. \ref{feyps}.

To achieve this let
\begin{equation} U=U(Q^{M}_{\neq 0}+Q^{M}_{0})=U_{\neq 0}+U_{0}.
\label{usplit}
\end{equation} Here
$Q^{M}_{0}=\sum_{j=3}^{\infty}P^{M}_{j}P_{0,j-2}+(P^{M}_{2}+P^{M}_{1})P_{0,1}$
is the projection operator for a lattice system in state
$|0\rangle$ being  at a lattice position two sites behind that of
the head M if $j\geq 3$ and at position $1$ if the head is at
sites $2$ or $1$. $Q^{M}_{\neq 0}$ is the projection operator for
the lattice system, at the same position relative to that of $M$,
in a symbol state different from $|0\rangle$.  Note that  the two
projection operators are orthogonal and $Q^{M}_{\neq
0}=1-Q^{M}_{0}$.  Also $[U_{\neq 0},U_{0}]\neq 0.$

The definition of $Q^{M}_{0}$  and $Q^{M}_{\neq 0}$ is based on
the observation that for any symbol $S$, including $0$,
$UP_{k}^{M}P_{S,j-2} = P_{S,j-2}UP_{k}^{M}$ for $k\geq j$ where
$P^{M}_{j}$ and $P_{S,j}$ are projection operators for $M$ at site
$j$ and the site $j$ lattice system in state $|S\rangle$. This
holds because the properties of $U$ are such that the state of any
lattice system located $2$ or more sites behind $M$ is out of
range and unchanged by the action of $U$.

Based on this  $U^{n}$ can be expanded into sums of products of
$U_{\neq 0}$ and $U_{0}$: \begin{equation} (U_{\neq 0}+U_{0})^{n}
= \sum_{\nu(1)=0,1}\sum_{t=1}^{n}\sum_{h_{1},\cdots
,h_{t}=1}^{\delta_{\Sigma
,n}}U_{\nu(t)}^{h_{t}}U_{\nu(t-1)}^{h_{t-1}}\cdots
U_{\nu(2)}^{h_{2}}U_{\nu(1)}^{h_{1}}. \label{feypsc}
\end{equation}
Here $\nu$, $t$, and $h$  have the same meaning as in Eq.
\ref{wdstrng}. The upper limit $\delta_{\Sigma,n}$ on the $h$ sums
expresses the condition that $\sum_{k=1}^{t}h_{k}=n$.

Eq. \ref{feypsc} can be used to expand $\Psi(n)$ into a sum over
word path states similar to the sum over symbol path states shown
in Eq. \ref{feyps}. From Eq. \ref{feypsa} one has
\begin{eqnarray*}U^{m}|\ell_{j+1},j+1,\underline{0}_{[j+1,j+m]},S_{j}^{\prime}\rangle = \\
\sum_{\ell_{j+m+1},S_{j+m}^{\prime}}\sum_{
\underline{S}_{[j,j+m-1]}}|\ell_{j+m+1},j+m+1,S_{j+m}^{\prime},
\underline{S}_{[j,j+m-1]}\rangle \times \\ \langle
\ell_{j+m+1},j+m+1,S_{j+m}^{\prime}, \underline{S}_{[j,j+m-1]}
|U^{m}|\ell_{j+1},j+1,\underline{0}_{[j+1,j+m]},S_{j}^{\prime}\rangle
.\end{eqnarray*} Use of this and completeness relations to remove
the intermediate sums over $M$ states and $S^{\prime}$ states,
gives

\begin{eqnarray} \Psi(n)=|\underline{0}_{[>n+1]}\rangle\otimes
\sum_{\nu(1)=0,1}
\sum_{\ell_{n},S_{n+1}^{\prime}}\sum_{t=1}^{n}\sum_{
\underline{p}} \sum_{h_{1},\cdots ,h_{t}=1}^{\delta_{ \Sigma ,n}} \nonumber \\
|\ell_{n},n+2,S_{n+1}^{\prime},\underline{p}\rangle
  \langle
\ell_{n},n+2,S_{n+1}^{\prime},\underline{p}(t)|U^{h_{t}}_{\nu
(t)}|\underline{0}_{[n+2-h_{t},n+1]}\rangle\times \nonumber \\
\cdots \langle
\underline{p}(2)|U^{h_{2}}_{\nu(2)}|\underline{0}_{[h_{1}+1,h_{1}+h_{2}]}\rangle\langle
\underline{p}(1)|U^{h_{1}}_{\nu(1)}|i,2,\underline{0}_{[1,h_{1}]}\rangle.
\label{wdpthsum}
\end{eqnarray}  The path sum is over all paths $\underline{p}$ containing $t$
words and a total of $n$ symbols.  This can also be expressed
using projection operators by \begin{eqnarray} \Psi(n) =
\sum_{\nu(1)=0,1} \sum_{t=1}^{n}\sum_{ \underline{p}}
\sum_{h_{1},\cdots ,h_{t}=1}^{\delta_{\Sigma,n}}P_{
\underline{p}}U_{\nu(t)}^{h_{t}}
P_{\underline{0}_{[n+2-h_{t},n+1]}}\times \nonumber \\
P_{\underline{p}(t-1)}U_{\nu(t-1)}^{h_{t-1}} \cdots
U_{\nu(2)}^{h_{2}}P_{ \underline{0}_{h_{1}+1,h_{1}+h_{2}}}P_{
\underline{p}(1)}U_{\nu(1)}^{h_{1}}|i,2,\underline{0}\rangle.
\label{pthexp} \end{eqnarray}  Here $P_{ \underline{p}(j)}$ and
$P_{ \underline{0}_{[a,b]}}$ are projection operators on the $jth$
word in path $ \underline{p}$ and on the spacer string over the
lattice interval $[a,b]$.  Each projection operator $P_{
\underline{p}(j)}$ commutes past all operators standing to the
left of it in the equation. The number of nonempty words in
$\underline{p}$ is $t/2$ if $t$ is even.  If $t$ is odd the number
of words is $(t-1)/2$ if $\nu(1) =0$ and $(t+1)/2$ if  $\nu(1)=1$.

The following definitions and properties of the mathematical
logical concepts for the quantum mechanical example are based on
the assignment of meaning to some of the word states in $|
\underline{p}\rangle$ where the meaning is based on the tree
structure of the paths shown in two equations. An informal
discussion of these concepts is combined with  more precise
definitions in terms of expectation values of projection
operators. Details, including proofs of the existence of the
limits involved, are given in \cite{BenDTVQM}.

From now on word states will be assumed to be nonempty (contain
no $0s$).  Also underlining will be suppressed. A word state
$|X\rangle$ is defined to be {\em printable} if it appears in
some path at some time. That is $|X\rangle$ is printable if
\begin{equation}\lim_{n\rightarrow\infty}(\Psi(n)|Q^{M}_{X}|\Psi(n))
>0.\label{xpr}\end{equation} $|X\rangle$ is not printable if the limit in Eq.
\ref{xpr} equals $0$. From now on $|X\rangle$ will be referred to
as a word.

Here \begin{equation} Q^{M}_{X}=
\sum_{a=1}^{\infty}Q^{M}_{X,a}=\sum_{a=1}^{\infty}\sum_{j=2}^{\infty}
P^{M}_{b+j}P_{0X0_{[a,b]}} \label{qproj}\end{equation} with
$b=a+L(X)+2$ where $L(X)$ is the number of symbols in $X$.
$Q^{M}_{X}$ is the projection operator for finding $|X\rangle$
followed and preceded by at least one $|0\rangle$ with
$|0_{b}\rangle$ located two or more sites behind M. $P^{M}_{b+j}$
and $P_{0X0_{[a,b]}}$ are projection operators for finding $M$ at
site $b+j$ and the word $|0X0\rangle$ starting at site $a$ and
ending at site $b$. The reason for the $0s$ before and after $X$
is to exclude cases where $X$ is part of a longer word.  The
requirement that the terminal $0$ in $|0X0\rangle$ be at least two
sites behind M means that the probability that $0X0$ appears at a
fixed location is independent of $n$ for sufficiently large $n$.
More exactly $(\Psi(n)|Q^{M}_{X,a}|\Psi(n))$ is independent of $n$
for all $n\geq b+2$.

The meanings chosen for the sentences $|P(X)\rangle$ and $|\sim
P(X)\rangle$ are based on the path description of Eq.
\ref{wdpthsum} or \ref{pthexp}. A sentence is a word that has a
meaning. The domain of meaning for $|P(X)\rangle$ is the set of
paths containing $|P(X)\rangle$. Similarly the meaning domain for
$|\sim P(X)\rangle$ is the set of paths containing $|\sim
P(X)\rangle$. $|P(X)\rangle$ is {\em true} on its domain if all
paths containing $|P(X)\rangle$ also contain $|X\rangle$. It is
false if some paths containing $|P(X)\rangle$ do not contain
$|X\rangle$. $|\sim P(X)\rangle$ is {\em true} on its domain if no
path containing $|\sim P(X)\rangle$ contains $|X\rangle$. It is
false if some path containing $|\sim P(X)\rangle$ contains
$|X\rangle$. Note that $|P(X)\rangle$ and $|X\rangle$ are distinct
words in a path, so they are separated by at least one spacer
string. Also the order of appearance of $|X\rangle$ and
$|P(X)\rangle$ in the path is immaterial.

These definitions can be expressed as limits of matrix elements.
$|P(X)\rangle$ is true if
\begin{eqnarray} \lim_{n,m\rightarrow\infty}\langle\Psi(n)|
Q^{M}_{P(X)}(U^{\dagger})^{m}Q^{M}_{X}U^{m}Q^{M}_{P(X)}|\Psi(n)\rangle
 \nonumber \\
=\lim_{n\rightarrow\infty}\langle\Psi(n)|Q^{M}_{P(X)}|\Psi(n)\rangle
\label{pxtr} \end{eqnarray} and $|\sim P(X)\rangle$ is true if
\begin{eqnarray}\lim_{n,m\rightarrow\infty}\langle\Psi(n)| Q^{M}_{\sim
P(X)}(U^{\dagger})^{m}Q^{M}_{\neg X}U^{m}Q^{M}_{\sim
P(X)}|\Psi(n)\rangle \nonumber \\ =\lim_{n\rightarrow\infty}
\langle\Psi(n)|Q^{M}_{\sim
P(X)}|\Psi(n)\rangle.\label{notpxtr}\end{eqnarray} Here
$Q^{M}_{\neg X}$ is the projection operator for $|X\rangle$ not
occurring in any path two or more sites behind the head. Note that
$Q^{M}_{X} =1-Q^{M}_{\neg X}$ inside these matrix elements.
$|P(X)\rangle$ and $|\sim P(X)\rangle$ are false if these
equations hold with $=$ replaced by $<$.

These equations show that if a measurement at time $n$ of
$Q_{P(X)}^{M}$ finds $P(X)$ then $|P(X)\rangle$ is true [false] if
the asymptotic conditional probability is unity [$<1$] that $X$
will be found in a subsequent measurement of $Q_{X}^{M}$.
Similarly if $\sim P(X)$ is found in a measurement of $Q_{\sim
P(X)}^{M}$, then $|\sim P(X)\rangle$ is true [false] if the
asymptotic conditional probability that $X$ will not be found is
unity [$<1$].

Based on the choice of meaning used here, $|P(X)\rangle$ says
nothing about the occurrence or nonoccurrence of $|X\rangle$ in
paths not containing the sentence.  The same holds for $|\sim
P(X)\rangle$. For a measurement at time $n$ that does not find
$P(X)$, the truth or falseness of $|P(X)\rangle$ has no meaning
for any following measurement of $|X\rangle$, relative to the time
$n$ measurement. However a later measurement may find $P(X)$. This
shows that the meaning domain is nondecreasing with increasing
$n$.  One concludes from this that the domain of meaninglessness
of $|P(X)\rangle$ is not empty if and only if
$\lim_{n\rightarrow\infty}\|Q^{M}_{\neg P(X)}\Psi(n)\|>0.$  A
similar statement holds for $|\sim P(X)\rangle$.

The definitions of truth and falseness  given above  relate these
concepts to the dynamics $U$. But nothing so far requires the
sentences to be true.  This is the case even if $U$ is a correct
theoretical description of the dynamics of $M$.

This is taken care of by defining the dynamics $U$ to be {\em
valid} if for all paths $|p\rangle$ and all sentences $|S\rangle$,
if $|p\rangle$ contains $|S\rangle$ then $|S\rangle$ is true in
$|p\rangle$. An equivalent statement is that $U$ is valid if all
printable sentences are true on their domain of meaning. In terms
of limits of matrix elements one has that $U$ is valid if for all
sentences $|S\rangle$ for which Eq. \ref{xpr} holds, so do Eqs.
\ref{pxtr} for $|S\rangle=|P(X)\rangle$ and \ref{notpxtr} for
$|S\rangle=|\sim P(X)\rangle$.

$U$ is {\em consistent} if for all $|X\rangle$ that are not
sentences, no path contains both $|\sim P(X)\rangle$ and
$|P(X)\rangle$. In terms of limits of matrix elements, $U$ is
consistent if \begin{equation} \lim_{n,m\rightarrow\infty}\langle
\Psi(n)|Q^{M}_{P(X)}(U^{\dagger})^{m} Q^{M}_{\sim P(
X)}U^{m}Q^{M}_{P(X)}|\Psi(n)\rangle
=0.\label{consis}\end{equation}

The definition of validity has some interesting aspects. It is
satisfying to note that, as is the case for the classical $M$, one
can prove that if $U$ is valid, then it is consistent.  The
converse does not necessarily hold, though.

The requirement that $U$ is valid is a restriction on  $U$ as it
limits what can and cannot appear in paths containing the
sentences. This requirement corresponds to conditions that must be
satisfied by the amplitudes associated with the paths in the path
sum, Eq. \ref{wdpthsum} or \ref{pthexp}.

One property of the definition  is that it says nothing about how
many, if any, of the words $|P(X)\rangle$ and $|\sim P(X)\rangle$
are printable by $U$. For instance $U$ is valid if no sentence is
printable. One would like to avoid this possibility. Also it is
desirable for $U$ to maximize the printing of  words that give
information about the dynamics by telling what can and cannot be
printed.

To this end one defines $U$ to be {\em complete} if all sentences
are printable. If $U$ is complete then, for the interpretation
considered here, the amount of information provided by $U$ about
what it can and cannot print is maximal. Whether or not this
condition can be satisfied, or should be relaxed in the presence
of conditions that exclude printing of some sentences, may depend
on the interpretation given to the words chosen to be sentences.
More generally one defines $U$ to be {\em maximally complete} if
all sentences that are not excluded by these conditions, if any,
are printable.

It should be noted that, for the path interpretation used here,
completeness is quite different for the quantum $M$ than for the
classical single path $M$. In particular completeness does not
restrict the printability of sentences and their negation. Both
$|P(X)\rangle$ and $|\sim P(X)\rangle$ can appear provided they
are on different paths. This ensures that consistency is
satisfied. This is impossible for a classical $M$ with only one
path as at most one of $|P(X)\rangle$ or $|\sim P(X)\rangle$ can
appear.

The relationship among these concepts is shown in the figure which
is a schematic tree representation of very few of the paths  in
the sum over word paths of Eqs. \ref{wdpthsum} or \ref{pthexp}.
For illustrative purposes, only a very few of the relevant words
are shown in some of the path segments. A full tree representation
would be very complex with 5 branches at each time step node. This
is based on the sum over symbol paths of Eq. \ref{feyps}.
\begin{figure}
\vspace{6cm} \caption{Tree representation of some of the word
paths in the path sum.  All paths have the same length of $n$
symbols and grow upwards at the same rate. The ordinate location
of a word by a path segment denotes approximately when the word
appeared in that path segment. Each word is separated from an
adjacent word by one or more $0$s.}
\end{figure}

The paths all have the same height as they each contain $n$
symbols.  Words next to the paths denote that the paths contain
these words.  The two paths with circled words show examples of
invalidity.  The path containing $|P(W)\rangle$ and $|\sim
P(W)\rangle$ is inconsistent and the path containing $|X\rangle$
and $|\sim P(X)\rangle$ shows that $|\sim P(X)\rangle$ is false.
The validity status of the path containing $|P(X)\rangle$ and no
$|X\rangle$ is still open as $|X\rangle$ may appear later on.

\section{Dependence of Validity on the Basis}
\label{DVB}  It should be emphasized that the word states chosen
to have meaning and the meaning assigned to these states were
chosen or imposed arbitrarily from the outside.  They were done
to illustrate properties of some mathematical logical concepts in
quantum mechanics.   In particular the requirement that the
dynamics of $M$ be valid imposes a restriction on the dynamics
that depends on the meaning assigned to the states and to the
truth definitions used.  The quantum system $M$ is itself
completely silent on which expressions, if any, have meaning and
how they are to be interpreted. $M$ is also silent on what basis
is to be used to to assign meaning to the states in the basis.

In this connection it is worth investigating the dependence of
validity of a dynamics $U$ on a change of basis. Intuitively one
would expect that validity of a dynamics would not be preserved
under a change of basis.  Reading a word state in different basis
would give a different outcome with a finite probability that
depends on the relationship between the two basis.  Also, as is
well known, the state is changed by the reading so that it cannot
be reread to determine the outcome in the original basis.

As a simple example of this, suppose $M$ is generating output as
described earlier in Section \ref{QMM}. Assume that the output
lattice is a lattice of spin $2$ systems with the spin projection
eigenstates of each system along some axis corresponding to the 5
symbol states. Initially each system is in a spin projection
eigenstate corresponding to the $|0\rangle$ symbol state.   Then
meaningful output word states $|W_{[a,b]}\rangle$, i. e. those of
the form $|P(X)\rangle$ or $\sim |P(X)\rangle$ where $|X\rangle$
is any word that is not a sentence, correspond to products of spin
projection states. Also the truth, validity, and completeness of
the dynamics of M is defined in terms of these states.

Suppose that the dynamics of $M$ is the same but the basis, chosen
by some observer $O$, for reading the output of $M$ is changed.
For example, assume that $O$ uses a different axis for observing
the spin projection states but keeps the same correspondence
between spin projections and symbols. In this case one expects
that any dynamics $U$ which is valid for the original basis is
not, in general, valid for the new basis used by $O$.

To see this let $U(\Omega)$ be the unitary rotation operator that
maps states in the original basis to those in the rotated basis
where $\Omega$ is the rotation angle between the two axes. The
amplitude for finding expression $|X_{[c,d]}\rangle$ in the
rotated basis $m$ steps after finding $|\sim P(X)_{[a,b]}\rangle$
at time step $n$ also in the rotated basis is given by
\begin{eqnarray} |\sum_{W,Y,Z}\langle
X_{[c,d]}|U(\Omega)|Z_{[c,d]}\rangle\langle
Z_{[c,d]}|U^{m}|Y_{[a,b]}\rangle\langle
Y_{[a,b]}|U(\Omega)^{\dagger}|\sim P(X)_{[a,b]}\rangle \otimes &
\nonumber \\ \langle \sim
P(X)_{[a,b]}|U(\Omega)|W_{[a,b]}\rangle\langle
W_{[a,b]}|U^{n}|i,2,\underline{0}\rangle |. & \end{eqnarray}  In
this amplitude, which is based on Eqs. \ref{feyps}, \ref{feypsb},
and \ref{qproj}, the sums over the $M$ states and the primed
symbol states (Eqs. \ref{feyps} and \ref{feypsb}) are suppressed
as are the two $0s$, one before and one after the expressions. The
$W,Y,Z$ sums are over all symbol string states in the original
basis of length $d-c+1$ for $Z$ and $b-a+1$ for $Y,W$ where the
$0$ symbol can occur in the string. The interval lattice locations
of these string states are shown by the subscripts where $b =a
+L(X)+6$ and $d=c+L(X)+2$. Also $n>b+2$ and $m+n>d+2$ with $L(X)$
equal to the length of $X$.

It is clear that this amplitude is not zero in general.  This
shows that $U$ is not valid in the rotated basis because it gives
a nonzero amplitude for both $X$ and $\sim P(X)$ to appear in a
path in the rotated basis at the specified lattice locations. This
holds even if $U$ is valid in the original basis where the matrix
element $\langle Z_{[c,d]}|U^{m}|Y_{[a,b]}\rangle =0$ for all $m$
if $|Y_{[a,b]}\rangle = | \sim P(X)_{[a,b]}\rangle$ and
$|Z_{[c,d]}\rangle =|X_{[c,d]}\rangle$ (original basis).  In this
case the other terms in the sums give the nonzero contributions
for the amplitude.

More generally, let $u$ be any unitary operator on the five
dimensional Hilbert space spanned by the symbol basis and let $u$
be independent of the lattice site. Define the  symbol projection
operators for the observer in terms of the original symbol
projection operators by \begin{equation}P^{O}_{S,j}
=uP_{S,j}u^{\dagger}.\label{proju}\end{equation} Here
$P^{O}_{S,j}$ and $P_{S,j}$ are projection operators for the
symbol $S$ at site $j$ in the observer reading basis and in the
original basis. The corresponding projection operators for words
in the observer basis and the original basis are obtained as
tensor products of these operators.

Define a new dynamics $V$ by \begin{equation}V=\omega
U\omega^{\dagger}\label{Vuu}\end{equation} where $\omega =
\sum_{j=2}^{\infty}P_{j}^{M}u_{j}\otimes u_{j-1}$. Here
$P^{M}_{j}$ is the projection operator for $M$ at site $j$ and
$u_{j}$ is the operator $u$ restricted to symbols at site $j$ of
the lattice. It is clear that $\omega^{\dagger}\omega =
\sum_{j=2}^{\infty}P_{j}^{M}=\omega\omega^{\dagger}$ is unitary on
the subspace of all states with $M$ at positions $\geq 2$.  Since
this is the space of states attained by iteration of $U$ on any
initial state with $M$ at sites $j\geq 2$ one can consider
$\omega$ to be unitary.

Define the projection operator $Q^{M,O}_{X}$ by replacing
$P_{0X0_{[a,b]}}$ by $P^{O}_{0X0_{[a,b]}}$ in Eq. \ref{qproj}
where $P^{O}_{0X0_{[a,b]}}$ is equal to the tensor product of
single symbol operators given by Eq. \ref{proju} over the lattice
site interval $[a,b]$. Replacement of $U$ by $V$ and each $Q^{M}$
operator by $Q^{M,O}$  in  Eqs. \ref{pxtr} and \ref{notpxtr} shows
that if $U$ is valid then so is $V$ but for a different initial
state $\omega |i,2,\underline{0}\rangle = |i,2\rangle\otimes
u_{2}|0_{2}\rangle \otimes u_{1}|0_{1}\rangle \otimes
|\underline{0}_{[>2]}\rangle .$ Furthermore the dynamics $V$ is
the same as $U$ if and only if $U$ and $\omega$ commute.

This shows that validity is  preserved  under a unitary change in
basis if and only if the unitary operator $u$ generating the
basis change commutes with the dynamics $U$.  Since this is not
the case in general one sees that validity is not preserved, in
general, under a change of basis.

\section{Incompleteness}
\label{GI} The quantum mechanical  model described in this paper
can be extended to show results similar to those expressed by the
G\"{o}del incompleteness theorems.  To this end one needs to be
able to describe word states that refer to their own printability
and unprintability. Following Smullyan \cite{Smullyan}, the symbol
state $|N\rangle$ is added to the language. The word
$|N(X)\rangle$ denotes or refers to the word $|X(X)\rangle$. The
set of sentences is expanded to include words of the form
$|PN(X)\rangle$ and $|\sim PN(X)\rangle$ where $|X\rangle$ is any
expression. For the path interpretation $|PN(X)\rangle$ [$|\sim
PN(X)\rangle$] mean that all paths containing $|PN(X)\rangle$
[$|\sim PN(X)\rangle$] contain [do not contain] $|X(X)\rangle$.

Here it is useful to ignore the problems with inference chains
resulting from this expansion \cite{BenDTVQM} and concentrate on
just two words, $|PN(\sim PN)\rangle$ and $|\sim PN(\sim
PN)\rangle$. Based on the path interpretation, the word $|\sim
PN(\sim PN)\rangle$ is self referential in that it means that all
paths containing $|\sim PN(\sim PN)\rangle$ do not contain $|\sim
PN(\sim PN)\rangle$. Since this is a contradiction, one concludes
that this interpretation is not possible for this word.

An equivalent argument based on the truth and validity
definitions is as follows:  Assume that $|\sim PN(\sim PN)\rangle$
is printable. Then Eq. \ref{notpxtr} shows that this sentence is
false (substitute $Q^{M}_{\neg \sim PN(\sim PN)}$ for
$Q^{M}_{\neg X}$ in Eq. \ref{notpxtr} and use Eq. \ref{consis} to
see that Eq. \ref{notpxtr} becomes an inequality).  From this one
concludes either that $U$ is not valid for this sentence, or $U$
is valid and $|\sim PN(\sim PN)\rangle$ is not printable and
therefore meaningless, or it has a meaning different from that
based on the path interpretation.

A similar argument holds for $|PN(\sim PN)\rangle$. If this word
were printable and $U$ is valid then the truth of $|PN(\sim
PN)\rangle$ means that $|\sim PN(\sim PN)\rangle$ must appear in
all paths containing $|PN(\sim PN)\rangle$.  But this is not
possible as has been seen. So $|PN(\sim PN)\rangle$ is false on
all paths containing it. Thus it either means something else, or
it has no meaning at all, or $U$ is not valid for this sentence.
As Eq. \ref{pxtr} shows, one cannot conclude it is not printable
and false.

To see the relation to the G\"{o}del incompleteness theorem, let
printability be a stand-in or surrogate for provability in
axiomatizable mathematical systems.  Then if $U$ is required to be
valid for all printable sentences, the above shows two sentences,
$|PN(\sim PN)\rangle$ and its negation, that cannot be printable
and maintain their intended meaning.  This corresponds to the
G\"{o}del incompleteness theorem for axiomatizable systems
\cite{Smullyan,Godel} where the proof of the theorem consists in
exhibiting a sentence, that refers to its own unprovability, and
its negation that cannot be theorems.

This is an example of conditions that exclude the printing of some
sentences that would have meaning if they were printable. In this
case one requires that $U$ is maximally complete in that all
sentences, except the two noted above, are printable.

\section{Meaning and Algorithmic Complexity}
\label{MAC} At this point it is worth a brief digression to look
at the relation between the meaning, if any, of quantum states in
general and their algorithmic complexity.  The meaning of the
states can be quite different from that considered in this paper
and the symbol basis can consist of more (or less) than $5$
states. What will be shown is that, if there is any such relation,
it must be complex and not at all obvious.  The proof consists of
showing two different dynamics $U_{1}$ and $U_{2}$ for $M$ that
have about the same algorithmic complexity.  But the dynamics are
quite different in that the states generated by $U_{1}$ have
meaning and those generated by $U_{2}$ do not.

To this end let $U_{1}$ and $U_{2}$ be the unitary dynamics for
two quantum Turing machines, $QTM_{1}$ and $QTM_{2}$. These
machines move $M$ as a multistate head in either direction along a
tape or lattice of quantum systems. Details of the system such as
the use of a two tape system will be suppressed to focus on the
essentials.

It is further required that $QTM_{1}$ and $QTM_{2}$ are quantum
theorem proving machines.  That is if $Ax_{1}$ and $Ax_{2}$ are
two different sets of axioms and $T_{1}$ and $T_{2}$ are the
theories based on $Ax_{1}$ and $Ax_{2}$, then iteration of $U_{1}$
on an empty tape or lattice state $|\underline{0}\rangle$,
generates or enumerates the theorems of $T_{1}$ as a product of
word states $|\underline{W}\rangle =
\otimes_{j=1}^{L(\underline{}W)}|\underline{W}_{j}\rangle$ where
each word $|\underline{W}_{j}\rangle$ is a theorem of $T_{1}$.
Here $L( \underline{W})$, the number of words in
$|\underline{W}\rangle$, is dependent on the number of iterations
of $U_{1}$.  Similarly iteration of $U_{2}$ on
$|\underline{0}\rangle$ generates a product word state consisting
of theorems of $T_{2}$.

In terms of matrix element amplitudes the meaning of this
requirement is that  for large $m$,
$|\langle\underline{W}|(U_{i})^{m}|\underline{0}\rangle |$ as a
function of $\underline{W}$ is strongly peaked around word string
states $|\underline{W}\rangle$ where for $i=1,2$ each word state
in $|\underline{W}\rangle$ is a theorem of $T_{i}$. Sums over
other degrees of freedom needed to ensure the unitarity of $U_{i}$
are suppressed in the amplitude.

The literature definition of quantum algorithmic complexity in
terms of lengths of product qubit states \cite{Berth,Vitanyi,Gacs}
can be used to define quantum algorithmic complexities for $U_{1}$
and $U_{2}$. To this end one notes that $U_{1}$ and $U_{2}$ each
consist of two parts; one part uses the logical rules of deduction
to generate new word states as theorems from those already present
and the other part inserts axioms as word states into
$|\underline{W}\rangle$ on request from the deduction part. The
deductive part is the same for $U_{1}$ and $U_{2}$ but the axiom
parts depend on the sets $Ax_{1}$ and $Ax_{2}$. Note that $Ax_{1}$
and $Ax_{2}$ have the same logical axioms. They differ in having
different nonlogical axioms.

Let $U$ be a universal quantum Turing machine that simulates
$U_{1}$ and $U_{2}$.  As is well known \cite{BerVaz} such machines
exist.  Let $|\underline{Z}_{i}\rangle$ be the input qubit string
state such that $U$ acting on
$|\underline{Z}_{i},\underline{0}\rangle$ simulates to good
accuracy the action of $U_{i}$ on $|\underline{0}\rangle$ for
$i=1,2$.  The length of $|\underline{Z}_{i}\rangle$ is determined
by three components.  Two are the same for each value of $i$ and
one depends on $i$.  The $i$ independent components are both of
finite length and include a part that depends on $U$, independent
of whatever machine $U$ is simulating, and another part that
simulates the application of the logical deduction rules. The $i$
dependent part is a program for generating the axioms in $Ax_{i}$.
This part is finite in length as it is decidable whether or not a
given word is or is not an axiom.

One now defines the quantum algorithmic complexity of $U_{1}$ and
$U_{2}$ to be the length of the shortest state
$|\underline{Z}_{i}\rangle$ such that $U$ acting on
$|\underline{Z}_{i},\underline{0}\rangle$  simulates to good
accuracy $U_{i}$ acting on $|\underline{0}\rangle$.  This extends
to quantum Turing machines the definition based on classical
machines \cite{Chaitin} that defines the algorithmic complexity of
a theory as the length of the shortest program as input to a
universal machine that generates the theorems of the theory.

This definition can now be used to complete the proof.  Let
$Ax_{1}$ be an axiom system that is consistent.  Use of one form
of the G\"{o}del completeness theorem \cite{Shoenfield}, that says
that an axiom system is consistent if and only if it has a model,
gives the conclusion that there exists an interpretation of the
theorems of $T_{1}$ into a model universe that gives them meaning.
Thus the word states generated by $U_{1}$ as theorems of $T_{1}$
have a meaning.

Let $Ax_{2}$ be obtained from $Ax_{1}$ by selecting one formula
$F$ from $Ax_{1}$ and adding its negation $\sim F$ to $Ax_{1}$.
Then $Ax_{2}$ contains both $F$ and its negation and all the other
axioms of $Ax_{1}$. It is clear that the algorithmic complexity of
$U_{2}$ for this case is essentially the same as for $U_{1}$. This
is the case because the increased length of
$|\underline{Z}_{2}\rangle$, which includes information needed to
carry out copying a formula $F$ and adding the negation symbol is
small compared to the length of $|\underline{Z}_{1}\rangle$.

However since $Ax_{2}$ is inconsistent, $T_{2}$ has no models, so
there is no interpretation of the theorem states generated by
$U_{2}$ that gives them meaning. This completes the proof.

This result shows that the  relation between the meaning of
quantum states and the algorithmic complexity of the dynamics that
generates the states must be quite complex, if there is any
relationship at all.  One hesitates to conclude that there is no
relationship at all because the above proof is based on an overall
framework or context that the word states have or do not have
meaning as theorems of axiomatizable theories. One must allow the
possibility that a different result might be obtained if the
output states of the $U_{i}$ were viewed in a different context.

\section{Discussion}

It should be noted that the correctness of $U$ for $M$ is
different from the validity of $U$ as used here.  $U$ is correct
for $M$ if calculated descriptions, based on the properties of
$U$, of the dynamical behavior of $M$ are correct.  This includes
calculations of the probability of occurrence of any word $X$ by
any time step $n$ and of other properties. $U$ is valid if some of
the words are assigned a meaning and these sentences are true on
their domain of meaning.  It is possible for $U$ to be correct and
not valid. This would be the case if $U$ correctly predicts the
(nonzero) probability of occurrence of a sentence that is false.

The path interpretation and resulting truth definitions for the
words $|P(X)\rangle$ and $|\sim P(X)\rangle$, Eqs. \ref{pxtr} and
\ref{notpxtr}, have the consequence that it is impossible for a
sentence to be not printable and false. This follows from the
fact that if the right hand limits of Eqs. \ref{pxtr} and
\ref{notpxtr} equal $0$, then these equations must hold as the
left hand limits are also $0$ as the matrix elements are all
nonnegative.

This supports the restriction of the meaning domain of a sentence
$|W\rangle$ to the paths containing $|W\rangle$. If $|W\rangle$
is not printable, it either has an empty meaning domain for the
intended interpretation or it has a different meaning  or
interpretation for which Eqs. \ref{pxtr} and \ref{notpxtr} do not
apply.

This limitation of meaning domains does not appear in some other
interpretations. For example, let printability be defined, as
before, by Eq. \ref{xpr}. Suppose $|P(X)\rangle$ is interpreted
to mean that $|X\rangle$ appears in any path, not just the paths
containing $|P(X)\rangle$, and $|\sim P(X)\rangle$is interpreted
to  mean that $|X\rangle$ appears in no paths at all. Then
$|P(X)\rangle$ is true if $\lim_{n\rightarrow\infty}\langle\Psi(n)
|Q^{M}_{X}|\Psi(n)\rangle >0$.  $|\sim P(X)\rangle$ is true if
$\lim_{n\rightarrow\infty}\langle\Psi(n)
|Q^{M}_{X}|\Psi(n)\rangle =0$.  In this case $|P(X)\rangle$ is
true if $|\sim P(X)\rangle$ is false and conversely, and there is
no restriction on the meaning domain of these sentences.  However
it is still the case that if these words are not printable then
$U$ does not provide information about its own dynamics in that it
says nothing about what can or cannot be printed.

As was noted already the choice of which states have meaning and
what these states mean was imposed externally.  $M$ was completely
silent about the meaning of its output.  As such this work is a
prelude to examining some much deeper and potentially more
interesting problems. Consider $M$ to be a complex quantum system,
such as a quantum robot, \cite{BenioffQR}, moving in and
interacting with a complex environment of quantum systems. As $M$
moves about it generates output or signals.  The state of the
cumulative output at time $t$ can be represented by a density
operator $\rho(t)$.  The time dependence of $\rho(t)$ allows for
the increase of the length or complexity of the output with
increasing $t$.  This increase with $t$ is the case for the
dynamics of M described by Eq. \ref{wdpthsum} where the length of
the word path states generated by M increases with the time step
number $n$. The density operator description is used to account
for the possibility that states of the output systems are
entangled with states of other quantum systems in M or in the
environment.

A basic question is "What properties must $\rho(t)$ have so that
we as external observers conclude that $\rho(t)$ has meaning?"
Even more important is the question "What properties must
$\rho(t)$ have so that we would conclude that it has meaning to M,
the system that generated it?"  And "Would the two meanings be the
same?"  If we interpreted $\rho(t)$ to be a theoretical and
experimental description of M's environment, and the
interpretation was valid, one might expect, and perhaps may even
require that $\rho(t)$ have the same meaning and interpretation
for M as for us as external observers.

In essence this problem is faced all the time by each human being
in interactions with other humans. All writing and speaking and
use of other means of communication can be described in terms of
some system M generating output that in essence creates systems
described by a time dependent state $\rho(t)$.\footnote{A specific
example of a quantum mechanical description of text as a
distribution of ink molecules on a space lattice is given in the
Appendix of \cite{BenioffTCTPM}.} The state is time dependent
because new output is being generated either continuously or
sporadically by M.  Each of us must be able to assign meaning to
the states of the output of others. This meaning is, for the most
part, the same as the meaning assigned by the system generating
the output.

A potentially important aspect of {\em existing} systems M that
generate output with meaning and of the output systems also is
that they are all large quantum systems for which the relevant
degrees of freedom are macroscopic or essentially classical. It is
suspected that that this may be a necessary condition.  The main
reason is that if the output consists of quantum systems in some
time dependent quantum state that is not quickly stabilized by
decoherent interactions with the environment \cite{Zurek}, then
reading the output to determine if it has meaning or not, requires
knowledge of what basis to use for the reading. As was shown in
Section \ref{DVB}, reading the output state in another basis will
give the wrong result with a finite probability that depends on
the relation between the two bases.  Also the state will be
changed so that one cannot read the state another time to get the
original answer.

\section{Summary and Conclusion}

In this paper an example of a machine $M$ generating output,
analyzed by Smullyan \cite{Smullyan} to illustrate various
mathematical logical concepts was described quantum mechanically.
Symbol strings states of the form $|P(X)\rangle$ and $|\sim
P(X)\rangle$ where $|X\rangle$ was any symbol string state without
$0s$ that did not have this form were assigned a path meaning. For
these words, referred to as sentences, $|P(X)\rangle$ was defined
to be true if all paths containing $|P(X)\rangle$ also contained
$|X\rangle$; $|\sim P(X)\rangle$ was true if no path containing
$|\sim P(X)\rangle$ also contained $|X\rangle$.

The dynamical evolution of $M$ described by iteration of a unitary
step operator $U$ was represented by a Feynman sum over word
paths. Based on this $U$  was defined to be valid if each sentence
was true on any path containing it. $U$ was defined to be complete
if each sentence appeared on some path, and $U$ was defined to be
consistent if  no path contained both $|P(X)\rangle$ and $|\sim
P(X)\rangle$.

Based on these definitions it was seen that the mathematical
logical concepts of truth, validity, completeness, and
consistency, have different properties than in the classical case.
For instance the domain of meaning of a sentence was limited to
the paths containing it. Sentences had no truth value for paths
not containing them.  Also, contrary to the classical case  for
which there is just one path, it is possible for $U$ to be valid
and consistent and for both a sentence and its negation to appear
on some paths. However no sentence and its negation can have any
path in common. It was also seen that a slight extension of the
model to include self referential sentences gives an
incompleteness result similar to that of the first G\"{o}del
incompleteness theorem.

A main purpose of this paper was to present and emphasize the main
results of \cite{BenDTVQM} without the extensive intervening
mathematics.  The figure was presented to illustrate more clearly
the above definitions and their relationships.  New material
includes showing that the properties defined above depend on the
basis used to define symbol states.  For instance it was seen that
validity of a dynamics $U$ was not preserved under a unitary
change of the symbol basis. However there is a transformed
dynamics that does preserve validity provided the initial state is
changed suitably. As was seen in the discussion this result is
relevant to the question regarding how one determines if output
generated by a quantum system $M$ moving in a quantum environment
has meaning, if any, to $M$. If the output system states are not
stabilized by interaction with the environment, one must know what
basis to  use to examine the output to answer this question.

Another new result was obtained by examining the relation between
the potential meaning of word string states in general and the
algorithmic complexity of the systems generating the word string
states.  Two word string states were described that had
essentially the same algorithmic complexity. For one string state
the component word states had meaning.  For the other they had no
meaning.  The contextual basis of the two states was the same in
that they were both theorem enumerations based on two different
axiom sets, one consistent and the other inconsistent. This shows
that the relationship between the potential meaning of a word
string state and the algorithmic complexity of the dynamics
generating the string must be quite complex, if any relationship
even exists.

In conclusion it is noted that the work done in this paper is a
small initial part of a larger attempt to combine mathematical
logical concepts with quantum mechanics.  This is one approach to
the questions presented in the discussion section, and towards the
goal of construction of a coherent theory of mathematics and
physics together.

\section*{Acknowledgements}
This work is supported by the U.S. Department of Energy, Nuclear
Physics Division, under contract W-31-109-ENG-38.

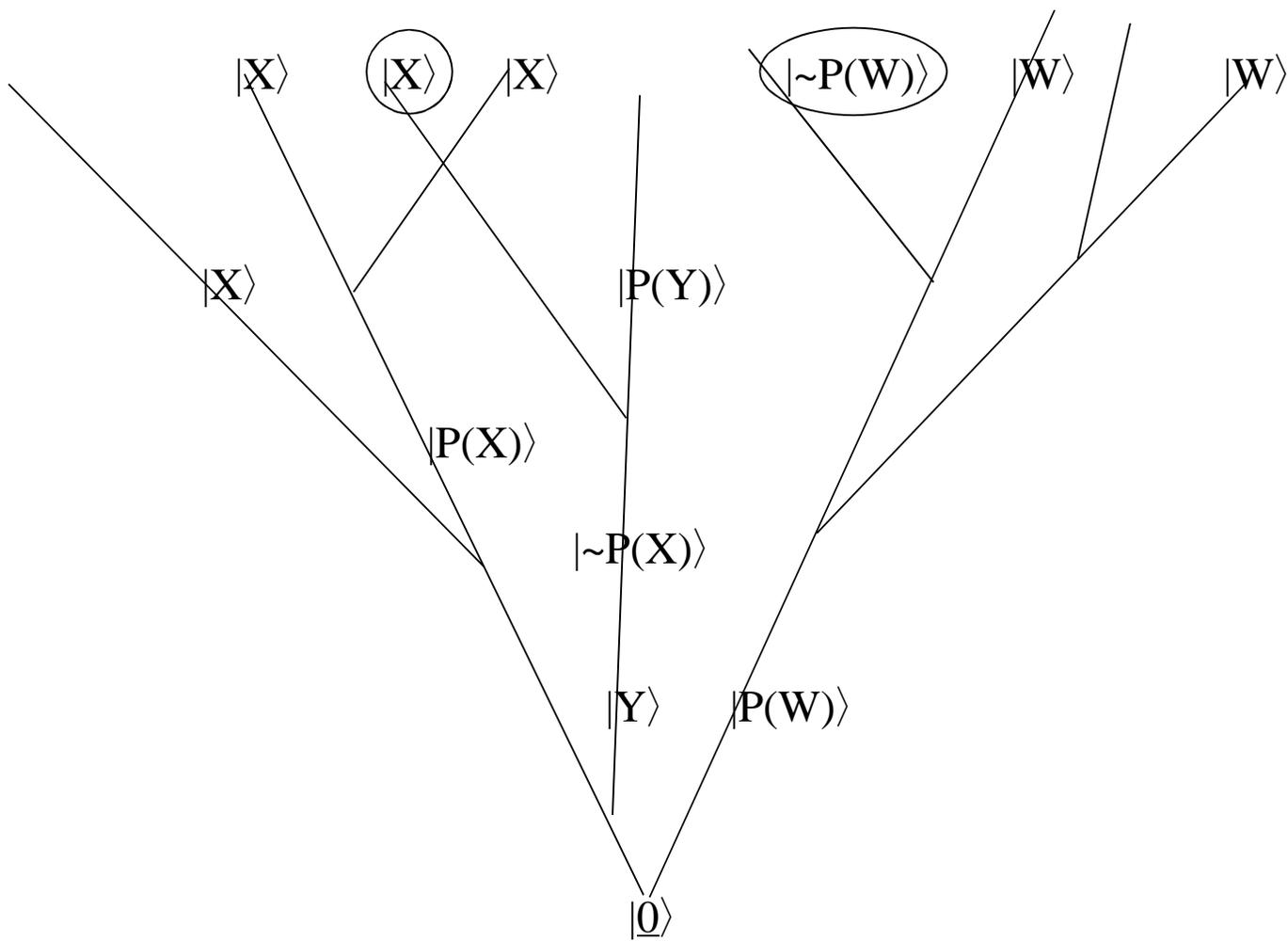
\end{document}